\documentstyle[multicol,aps,epsf]{revtex}
\begin{document}
\title{Finiteness and Fluctuations in Growing Networks}
\author{P.~L.~Krapivsky and S.~Redner}
\address{Center for BioDynamics, Center for Polymer Studies, 
and Department of Physics, Boston University, Boston, MA, 02215}

\maketitle
\begin{abstract}
  We study the role of finiteness and fluctuations about average quantities
  for basic structural properties of growing networks.  We first determine
  the exact degree distribution of finite networks by generating function
  approaches.  The resulting distributions exhibit an unusual finite-size
  scaling behavior and they are also sensitive to the initial conditions.  We
  argue that fluctuations in the number of nodes of degree $k$ become
  Gaussian for fixed degree as the size of the network diverges.  We also
  characterize the fluctuations between different realizations of the network
  in terms of higher moments of the degree distribution.


\end{abstract}
\begin{multicols}{2}
  
\section{Introduction}

Networks such as the Internet and the World-Wide Web do not grow in an
orderly manner.  For example, the Web is created by the uncoordinated effort
of millions of users and thus lacks an engineered architecture.  Although
such networks are complex in structure \cite{broder,willinger}, their large
size is a simplifying feature, and for infinitely large networks the rate
equation approach \cite{krrev} provides analytical predictions for basic
network characteristics.  Nevertheless, social and technological networks are
not large in a thermodynamic sense ({\it e.g.}, the number of molecules in a
glass of water vastly exceeds the number of routers in the Internet).  Thus
fluctuations in network properties can be expected to play a more prominent
role than in thermodynamic systems\cite{HA}.  Additionally, extreme
properties, such as the degree the node with the most links in a network
\cite{amaral,KR02}, the website with the most hyperlinks, or the wealth of
the richest person in a society, are important characteristics of finite
systems.  The size dependence of these properties or their distribution is
difficult to treat within a rate equation approach.

In this paper, we examine the role of finiteness and the nature of
fluctuations about mean values for large, but finite growing networks.  We
shall focus primarily on the degree distribution $N_k(N)$, the number of
nodes that are linked to $k$ other nodes in a network of $N$ links, as well
as related local structural characteristics.  We shall argue that self
averaging holds for the degree distribution, so that the random variables
$N_k(N)$ become sharply peaked about their average values in the $N\to\infty$
limit.  We shall also argue that the probability distribution for the number
of nodes of fixed degree, $P(N_k,N)$, is generally a Gaussian, with
fluctuations that vanish as $N\to\infty$.  On the other hand, higher moments
of the degree distribution do not self average.  This loss of self-averaging
ultimately stems from the power-law tail in the degree distribution itself.

In the next section, we define the growing network model and briefly review
the behavior of the average degree distribution in the thermodynamic
$N\to\infty$ limit.  We also discuss how the average degree distribution can
naturally be expected to attain a finite-size scaling form for large but
finite $N$.  We then describe our general strategy for studying fluctuations
in these growing networks.  In Sec.~III, we outline our simulational approach
and present data for the average degree distribution.  In the following two
sections, we examine the role of finiteness on the degree distribution, both
within a continuous formulation based on the rate equations (Sec.~IV), and an
exact discrete approach (Sec.~V).  The former approach is the one that is
conventionally applied to study the kinetics of evolving sustems, such as
growing networks.  While this approach has the advantage of simplicity and it
provides an accurate description for the degree distribution in an
appropriate degree range, it is quantitatively inaccurate in the large degree
limit.  This is the domain where discreteness effects play an important role
and the exact discrete recursion relations for the evolution of the degree
distribution are needed to fully account it properties.  In Sec.~VI, we
discuss the implications of our results for higher moments of the degree
distribution and their associated fluctuations.  Sec.~VII provides
conclusions and some perspectives.  Calculational details are given in the
appendices.

\section{Statement of the Problem}

The growing networks considered in this work are built by adding nodes to the
network one at a time according to the rule that each new node attaches to a
single previous node with a rate proportional to $A_k$, where $k$ is the
degree of the target node.  We investigate the class of models in which
$A_k=k+\lambda$, where $\lambda>-1$, but is otherwise arbitrary.  The general
situation of $-1<\lambda<\infty$, corresponds to linear preferential
attachment, but with an additive shift $\lambda$ in the rate.  This model was
originally introduced by Simon to account for the word frequency distribution
\cite{S}.  The case $\lambda=0$ corresponds to the Barab\'asi-Albert
model\cite{BA}, while the limit $\lambda\to\infty$ corresponds to random
attachment in which each node has an equal probability of attracting a
connection from the new node.  Thus by varying $\lambda$, we can tune the
relative importance of popularity in the attachment rate.

Previous work on the structure of such networks was primarily concerned with
the configuration-averaged degree distribution $\langle N_k(N)\rangle$, where
the angle brackets denote an average over all realizations of the growth
process for an ensemble of networks with the same initial condition.
Additionally, most studies focused on the tail region where $k$ is much
smaller than any other scale in the system.  For attachment rate
$A_k=k+\lambda$, this average degree distribution has a power-law
tail\cite{S,KR01},
\begin{equation}
\label{nkw}
\langle N_k(N)\rangle =N\,n_k, \quad {\rm with}\quad  n_k\propto
k^{-(3+\lambda)}
\end{equation}
as $ N\to\infty$.  In the specific case of $A_k=k$, the average degree
distribution explicitly is\cite{S,BA,KR01,KRL,DMS}
\begin{equation}
\label{nk}
\left\langle N_k(N)\right\rangle=N\,n_k, \quad {\rm with}\quad 
n_k={4\over k(k+1)(k+2)}\,.
\end{equation}

For finite $N$, however, the degree distribution must eventually deviate from
these predictions because the maximal degree cannot exceed $N$.  To establish
the range of applicability of Eqs.~(\ref{nkw}), we estimate the magnitude of
the largest degree in the network, $k_{\rm max}$ by the extreme statistics
criterion $\sum_{k\geq k_{\rm max}}\left\langle N_k(N)\right\rangle\approx
1$.  This yields $k_{\rm max}\propto N^{1/(2+\lambda)}$.  We therefore
anticipate that the average degree distribution will deviate from
Eq.~(\ref{nkw}) when $k$ becomes of the order of $k_{\rm max}$.  The
existence of a maximal degree also suggests that the average degree
distribution should attain a finite-size scaling form
\begin{equation}
\label{Nkscal}
\langle N_k(N)\rangle \simeq N n_k F(\xi), \qquad \xi={k/ k_{\rm max}}.
\end{equation}
Some aspects of these finite-size corrections were recently studied in
Refs.~\cite{ZM,KK,DMS2,burda}.  One basic result of our work is that we can
compute the scaling function explicitly.  We find that this function is
peaked for $k$ of the order of $k_{\rm max}$ and that it depends
substantially on the initial condition.  In contrast, the small-degree tail
of the distribution -- the reason why such networks were dubbed scale-free --
is independent of $N$ and the initial condition.

To study finite networks where fluctuations can be significant, we need a
stochastic approach rather than a deterministic rate equation formulation.
For finite $N$, the state of a network is generally characterized by the set
${\bf N}=\{N_1,N_2,\ldots\}$ that occurs with probability $P({\bf N})$.  The
network state ${\bf N}$ evolves by the following processes:
\begin{eqnarray}
\label{N12}
(N_1,N_2)&\to& (N_1,N_2+1),\nonumber\\
\label{N1k}
(N_1,N_k,N_{k+1})&\to&  (N_1+1,N_k-1,N_{k+1}+1).\nonumber
\end{eqnarray}
The first process corresponds to the new node attaching to an existing node
of degree 1; in this case, the number of nodes of degree 1 does not change
while the number of nodes of degree 2 increases by 1.  The second line
accounts for the new node attaching to a node of degree $k>1$.

From these processes, it is straightforward, in principle, to write the
master equation for the joint probability distribution $P({\bf N})$.  It
turns out that correlation functions of a given order are coupled only to
correlation functions of the same and lower orders.  Thus we do not need to
invoke factorization (as in kinetic theory) and we could, in principle, solve
for correlation functions recursively.  However, this would provide much more
information than is of practical interest.  Typically we are interested in
the degree distribution, or perhaps two-body correlations functions of the
form $\langle N_i\, N_j\rangle$.  Even though straightforward in principle,
it is difficult to compute even the two-point correlation functions $\langle
N_i\, N_j\rangle$ for general $i$ and $j$.  In this work, we shall restrict
ourselves to the specific (and simpler) examples of $\langle N_1^2\rangle$,
$\langle N_1\,N_2\rangle$, and $\langle N_2^2\rangle$.  We will use these
results to help characterize fluctuations in finite networks.

\section{Simulation Method and Data}

\begin{figure}
  \narrowtext \vskip 0in\hskip 0.4in 
  \epsfxsize=2.4in \epsfbox{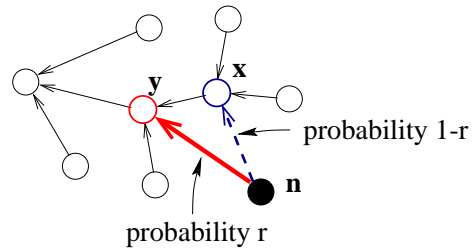}\vskip 0.1in

\caption{The re-direction process.  The new node {\bf n} selects a random
  target node {\bf x}.  With probability $1-r$ a link is established to this
  target node (dashed), while with probability $r$ the link is established to
  {\bf y}, the ancestor of {\bf x} (solid).
\label{R}}
\end{figure}

To simulate a network with attachment rate $A_k=k+\lambda$ efficiently, we
exploit an equivalence to the growing network with re-direction (GNR)
\cite{KR01}.  In the GNR, a newly-introduced node {\bf n} selects an earlier
``target'' node {\bf x} {\em uniformly}.  With probability $1-r$, a link from
{\bf n} to {\bf x} is created.  However, with probability $r$, the link is
{\em re-directed\/} to the ancestor node {\bf y} of node {\bf x}
(Fig.~\ref{R}).  As discussed in \cite{KR01}, the GNR is equivalent to a
growing network with attachment rate $A_k=k+\lambda$, with
$\lambda=r^{-1}-2$.  Thus, for example, the GNR with $r=1/2$ corresponds to
the growing network with linear preferential attachment, $A_k=k$.  Simulation
of the GNR is extremely simple because the selection of the initial target
node is purely random and the ensuing re-direction step is local.

There is, however, an important subtlety about this equivalence that was not
discussed previously in Ref.~\cite{KR01}.  Namely, the redirection process
does not apply when a node has no ancestor.  By construction, every node that
is added to the network does have a single ancestor, but some primordial
nodes may have none.  For example, for the very natural ``dimer'' initial
condition $\circ\!\!\!\longleftarrow\!\!\!\circ$, the seed node on the left
has no ancestor and the GNR construction for this node is ambiguous.  One way
to resolve this dilemma is to adopt the ``triangle'' initial condition in
which there are 3 nodes in a triangle with cyclic connections between nodes.
This leads to the correct attachment rate for each node for any value of
$\lambda$.  We therefore typically use this initial state to generate degree
distribution data.  On the other hand, theoretical analysis is simpler for
the dimer initial condition.  This state can also be simulated in a simple
manner (for the case $\lambda=0$) by a slightly modified GNR construction in
which direct attachment to the seed node is not allowed.  It is
straightforward to check that this additional rule leads to the correct
attachment rates for all the nodes in the network.

\begin{figure}
  \narrowtext \hskip 0.0in \epsfxsize=3.2in \epsfbox{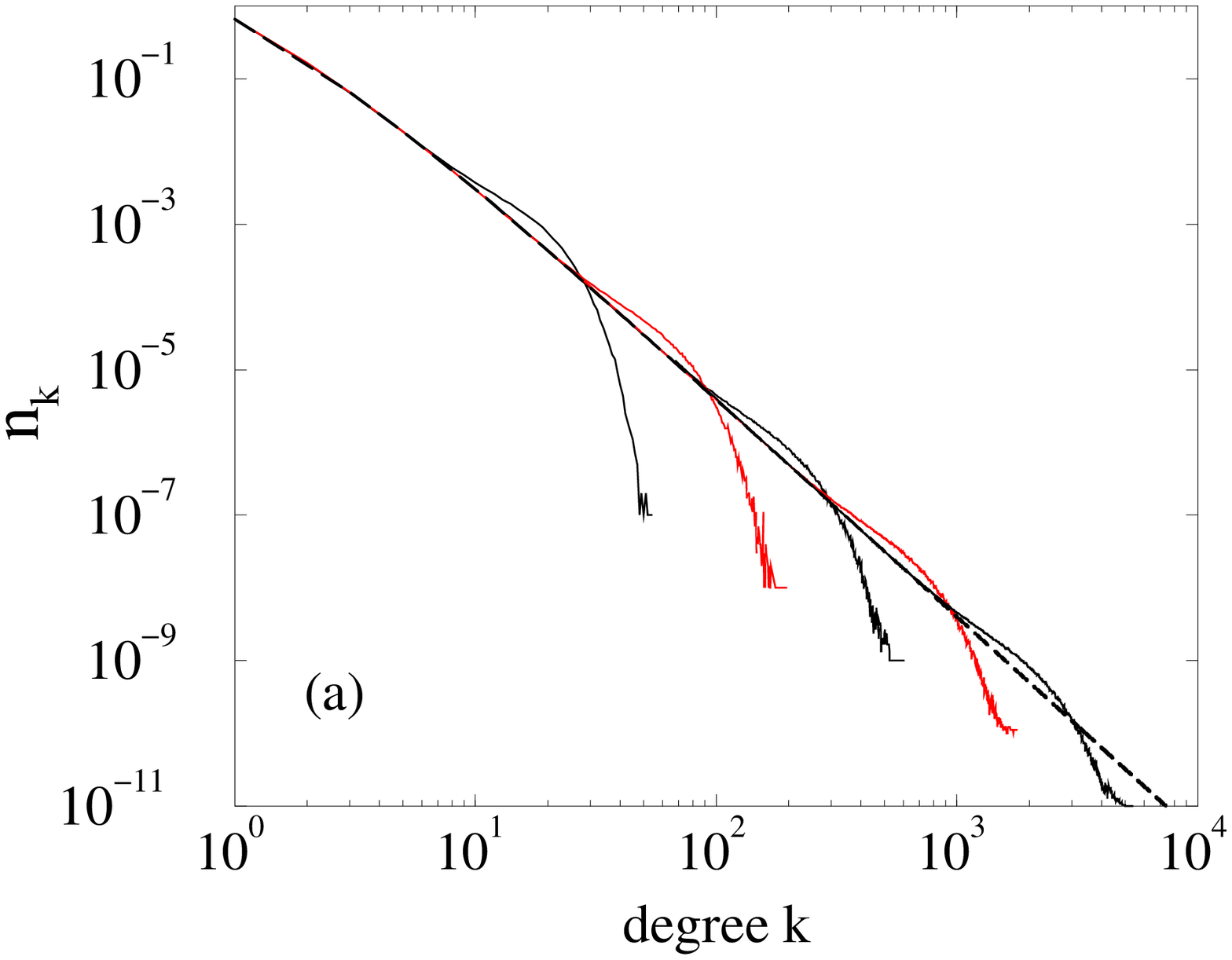} 
\vskip 0.1in \hskip 0.0in \epsfxsize=3.2in \epsfbox{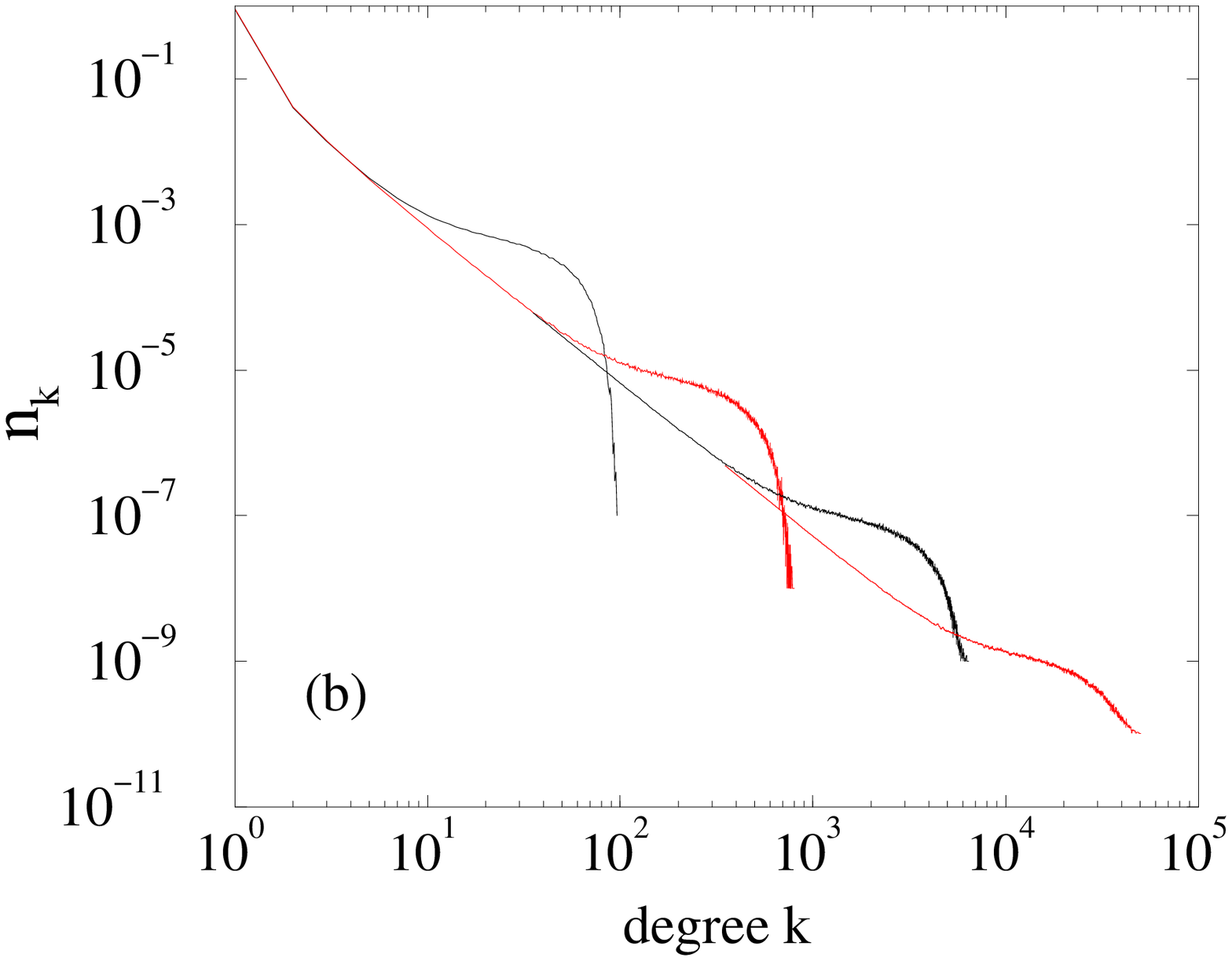} 
\caption{Normalized degree distributions for the triangle initial condition 
  for networks of $10^2,10^3,\ldots$ links (upper left to lower right), with
  $10^5$ realizations for each $N$, for (a) $A_k=k$ (up to $10^6$ links) and
  (b) $A_k=k+\lambda$, with $\lambda=-0.9$ (up to $10^5$ links).  In (a), the
  dashed line is the asymptotic result $n_k=4/[k(k+1)(k+2)]$; the last three
  data sets were averaged over 3, 9, and 27 points, respectively.  In (b),
  the last two data sets were averaged over 10 and 100 points, respectively.
\label{degree}}
\end{figure}

\begin{figure}
  \narrowtext \hskip 0.0in \epsfxsize=3.0in 
\epsfbox{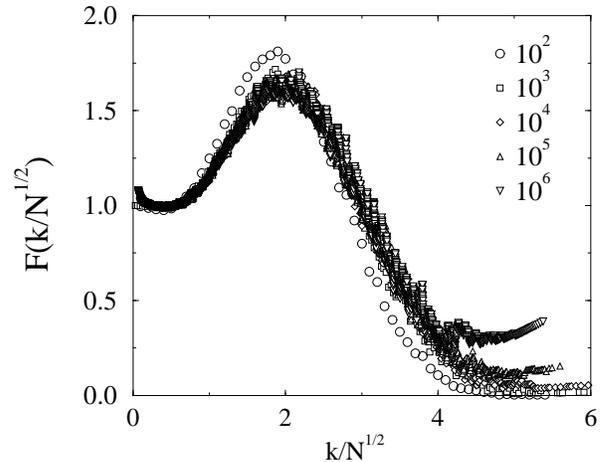}
\caption{The corresponding scaling function $F(\xi)$ in Eq.~(\ref{Nkscal}) for
  the data in Fig.~\ref{degree}(a).
\label{degree-scaling}}
\end{figure}

Figure~\ref{degree} shows the average degree distribution for attachment
rates $A_k=k$ and $A_k=k+\lambda$ with $\lambda=-0.9$ for the triangle
initial condition.  This latter value of $\lambda$ gives results that are
representative for values of $\lambda$ close to $-1$.  The data exhibits a
shoulder at $k\approx k_{\rm max}$ that is much more pronounced when
$\lambda<0$ (Fig.~\ref{degree}(b)).  This shoulder is also at odds with the
natural expectation that the average degree distribution should exhibit a
monotonic cutoff when $k$ becomes of the order of $k_{\rm max}$.  This
shoulder turns into a clearly-resolved peak that exhibits relatively good
data collapse when the degree distribution is re-expressed in the scaling
form of Eq.~(\ref{Nkscal}) (Fig.~\ref{degree-scaling}).  Conversely, the
magnitude of the peak diminishes rapidly when $\lambda$ is positive and
becomes imperceptible for $\lambda\agt 0.5$.

In the following two sections, we will attempt to understand this anomalous
feature of the degree distribution by studying the rate equations for the
node degrees of finite networks.

\section{Continuum Formulation}

We focus on the case of the linear attachment rate $A_k=k$ and briefly quote
corresponding results for other attachment rates.  In the continuum approach,
$N$ is treated as continuously varying.  Then the change in the average
degree distribution satisfies the rate equation
\begin{equation}
\label{NkN}
{d \left\langle N_k(N)\right\rangle\over dN}
=\left\langle {(k-1)N_{k-1}(N)
-k N_k(N)\over 2N}\right\rangle+\delta_{k,1}.
\end{equation}
We assume the dimer initial state -- two nodes connected by a single link so
that $N_k(N=1)=2\delta_{k,1}$.

Equations (\ref{NkN}) are recursive and can be solved sequentially, starting
with $\langle N_1\rangle$.  Explicit results for $\langle N_k\rangle$, $k\leq
4$, are given in Appendix A.  These expressions show that the dominant
contribution in the $N\to\infty$ limit is linear in $N$ and this corresponds
to the solution in Eq.~(\ref{nk}).  Indeed, if we substitute $\left\langle
  N_k(N)\right\rangle=n_k N$ into Eqs.~(\ref{NkN}), we obtain the recursion
$n_k=n_{k-1} (k-1)/(k+2)$, whose solution is Eq.~(\ref{nk}).  {}From the
first few $\langle N_k\rangle$, it is easy to see that the first correction
to this leading behavior is of the order of $N^{-1/2}$.  Substituting
$\left\langle N_k(N)\right\rangle=n_kN+A_k N^{-1/2}$ into Eqs.~(\ref{NkN})
and keeping the first two terms in each $\langle N_k\rangle$, we find
$A_k=4/3$.  Continuing this procedure systematically, we arrive at the
expansion:
\begin{eqnarray}
\label{nkn}
\left\langle N_k(N)\right\rangle
&=&n_k N+{4\over 3}\,{1\over N^{1/2}}-{3\over 2}\,{k-1\over N}\nonumber\\
&+&{4\over 5}\,{(k-1)(k-2)\over N^{3/2}}\nonumber\\
&-&{5\over 18}\,{(k-1)(k-2)(k-3)\over N^{2}}\nonumber\\
&+&{1\over 14}\,{(k-1)(k-2)(k-3)(k-4)\over N^{5/2}}+\ldots
\end{eqnarray}
In general, the right-hand side contains $k+1$ terms which can be written
more succinctly as
\begin{equation}
\label{expansion}
\left\langle N_k(N)\right\rangle=n_k N
+{1\over N^{1/2}}\sum_{j=0}^{k-1} {\Gamma(k)\over \Gamma(k-j)}\,\,
{(-1)^j \nu_j\over N^{j/2}}.
\end{equation}
The coefficients $\nu_j=(2j+4)/[j!(j+3)]$ may be obtained by imposing the
initial condition $N_k(1)=2\delta_{k,1}$ as each $\langle N_k\rangle$ is
computed; a simpler way of obtaining these coefficients will be explained
below.  Note that expansion (\ref{nkn}) is asymptotic because successive
terms decrease only when $k\ll \sqrt{N}$.

A more convenient way to solve Eqs.~(\ref{NkN}) is in terms of the generating
function
\begin{equation}
\label{genN}
{\mathcal N}(N,z)=\sum_{k=1}^\infty \left\langle N_k(N)\right\rangle\, z^k.
\end{equation}
Multiplying Eq.~(\ref{NkN}) by $z^k$ and summing over $k$, the generating
function satisfies the following partial differential equation
\begin{equation}
\label{gN}
\left(2N\,{\partial\over \partial N}
+z(1-z)\,{\partial\over \partial z}\right)
{\mathcal N}(N,z)=2Nz.
\end{equation}
The initial condition is ${\mathcal N}(1,z)=2z$, corresponding to a starting
point of two nodes and a single connecting link.

We reduce Eq.~(\ref{gN}) to a wave equation with constant coefficients by
changing from the variables $(N,z)$ to $(\ln\sqrt{N},\ln[z/(1-z)])$.  Then by
introducing the rotated coordinates $x, y$ such that $x+y=\ln\sqrt{N}$ and
$x-y=\ln[z/(1-z)]$, we recast the wave equation into
\begin{equation}
\label{gN1}
{\partial {\mathcal N}(x,y)\over \partial x}
={2\, e^{3x+2y}\over e^x+e^y}\,,
\end{equation}
whose general solution is
\begin{eqnarray*}
{\mathcal N}(x,y)=e^{2x+2y}-2e^{x+3y}+2e^{4y}\ln\left(e^x+e^y\right)+G(y).
\end{eqnarray*}
Finally, $G(y)$ is found by imposing the initial condition ${\mathcal
  N}(1,z)=2z$.  When $N=1$, we have $x=-y$, so that the initial condition
becomes ${\mathcal N}(-y,y)=2/(1+e^{2y})$.  Therefore
\begin{eqnarray*}
G(y)={2\over 1+e^{2y}}-1+2e^{2y} -2e^{4y}\ln \left(e^{-y}+e^y\right)
\end{eqnarray*}
and finally
\begin{eqnarray}
\label{nxy} 
{\mathcal N}(x,y)&=&e^{2y}\left(e^{2x}-2e^{x+y}+2\right)
+{1-e^{2y}\over 1+e^{2y}}\nonumber\\
&+&2e^{4y}\ln\left({e^{x+y}+e^{2y}\over 1+e^{2y}}\right).
\end{eqnarray}
Using $e^{2y}=(z^{-1}-1)\sqrt{N}$ and $e^{x+y}=\sqrt{N}$,
we re-express the generating function in term of the original variables 
\begin{eqnarray}
\label{exact}
{\mathcal N}(N,z)&=&(3-2z^{-1})N+2(z^{-1}-1)\sqrt{N}\nonumber\\
&+&{1-(z^{-1}-1)\sqrt{N}\over 1+(z^{-1}-1)\sqrt{N}}\nonumber\\
&-&2(z^{-1}-1)^2 N\,
\ln\left(1-z+{z\over \sqrt{N}}\right).
\end{eqnarray}

We are primarily interested in the degree distribution for nodes whose degree
is of the order of $k_{\rm max}\approx\sqrt{N}$.  This part of the
distribution can be extracted from the limiting behavior of the generating
function ${\mathcal N}(N,z)$ as $z\to 1$ from below.  Since the interesting
range is $k\approx\sqrt{N}$, it is convenient to write
\begin{equation}
\label{s}
z^{-1}=1+ {s\over \sqrt{N}}
\end{equation}
and keep $s$ finite while taking $N\to \infty$ limit.  We simplify still
further by eliminating the contribution to the generating function from the
power-law tail of $n_k$ in Eq.~(\ref{nk}).  For this purpose we consider the
modified generating function
\begin{equation}
\label{gen3N}
\left(z^2\,{\partial \over \partial z}\right)^3{\mathcal N}
=\sum_{k=1}^\infty (k+2)(k+1)k\left\langle N_k\right\rangle\, z^{k+3}
\end{equation}
which is constructed so that the derivatives multiply the degree distribution
by just the right factors to eliminate the power law tail.  The leading
behavior of this modified generating function will therefore provide the
scaling function $F(\xi)$ of Eq.~(\ref{Nkscal}).

We now substitute Eq.~(\ref{s}) and the anticipated scaling form of
Eq.~(\ref{Nkscal}) into the right-hand side of Eq.~(\ref{gen3N}) and replace
the sum by an integral.  This gives the Laplace transform of the scaling
function times a prefactor,
\begin{equation}
\label{LapF} 
4N^{3/2}\int_0^\infty d\xi\,F(\xi)\,e^{-\xi s}\,, 
\end{equation}
with $\xi=k/N^{1/2}$.  Using Eq.~(\ref{exact}), we compute the derivative on
the left-hand side of Eq.~(\ref{gen3N}).  In the $N\to \infty$ limit, this
derivative becomes $4N^{3/2} J(s)$ with  
\begin{equation}
\label{Js} 
J(s)={1\over 1+s} + {1\over (1+s)^2} 
+{1\over (1+s)^3} + {3\over (1+s)^4}\,. 
\end{equation}
This is just the Laplace transform of the scaling function.  Inverting the
Laplace transform then yields
\begin{equation}
\label{Fscal}
F(\xi)=\left(1+\xi\right)\left(1+{\xi^2\over 2}\right)\,e^{-\xi}\,.
\end{equation}
Notice that the coefficients $\nu_j$ in Eq.~(\ref{expansion}) can be obtained
by expanding $F$ in a Taylor series.  This is a much simpler approach than
solving each $\langle N_k(N)\rangle$ directly.

An important feature of the degree distribution is that it depends
significantly on the initial condition.  For example, for the triangle
initial condition, solving Eq.~(\ref{gN}) subject to
$N_k^\Delta(N=3)=3\delta_{k,2}$, or ${\mathcal N}^\Delta(3,z)=3z^2$, yields
\begin{eqnarray}
\label{exact2}
{\mathcal N}^\Delta(N,z)
&=&(3-2z^{-1})N+2(z^{-1}-1)\sqrt{3N}\nonumber\\
&+&3\left(1+(z^{-1}-1)\sqrt{N/3}\right)^{-2}-3\nonumber\\
&-&2(z^{-1}-1)^2 N\,
\ln\left(1-z+z\sqrt{3\over N}\right).
\end{eqnarray}
Repeating the steps used to deduce the scaling function (\ref{Fscal}) from
Eq.~(\ref{exact}), we now find
\begin{equation}
\label{Fscal2}
F^\Delta(\xi)
=\left(1+\eta+{\eta^2\over 2}+{\eta^4\over 4}\right)\,e^{-\eta},
\qquad \eta\equiv \xi\sqrt{3}. 
\end{equation}
Therefore small differences in the initial condition translate to
discrepancies of the order of $\sqrt{N}$ in the degree distribution of a
finite network of $N$ links.  Thus the properties of the nodes with the
largest degrees are quite sensitive to the first few growth steps of the
network (see also Ref.~\cite{KR02}).

While this initial condition dependence is real, there is also a spurious
aspect to this effect.  This may be illustrated by considering the linear
trimer initial condition $\circ$---$\circ$---$\circ$.  This is the unique
outcome of the dimer initial condition after one node has been added.  These
two initial conditions should therefore lead to the same degree distribution.
However, for the linear trimer initial state
($N_k(N=2)=2\delta_{k,1}+\delta_{k,2}$) the continuum approach gives the
scaling function,
\begin{eqnarray*} 
F(\xi)=\left(1+\eta+{\eta^2\over 2}+{\eta^3\over 4}
+{\eta^4\over 8}\right)\,e^{-\eta},
\qquad \eta\equiv \xi\sqrt{2}, 
\end{eqnarray*}
which is distinct from Eq.~(\ref{Fscal})!  This anomaly highlights one basic
limitation of the continuum formulation.

Finally, we mention that parallel results can be obtained for the general
case of the shifted linear attachment rate, $A_k=k+\lambda$.  The rate
equation for the average degree distribution is
\begin{eqnarray*}
{d \left\langle N_k(N)\right\rangle\over dN}
=\left\langle {A_{k-1}N_{k-1}(N)
-A_k N_k(N)\over A}\right\rangle+\delta_{k,1}\,,
\end{eqnarray*}
where $A=\sum A_k N_k= \sum (k+\lambda)N_k$.  To compute $A$ we use the sum
rules $\sum kN_k=2N$ (every link contributes twice to the total degree), as
well as $\sum N_k=N+1$ (for any tree initial condition) or $\sum N_k=N$ (for
an initial condition that has the topology of a single cycle).  To simplify
final formulae, we use the latter topology (specifically, the triangle
initial condition) so that $A=(2+\lambda)N$.

Solving the above rate equations successively, we find that the first two
terms in the asymptotic series for $\langle N_k^\Delta(N)\rangle$ are
\begin{equation}
\label{NkAw-soln}
\left\langle N_k^\Delta(N)\right\rangle\sim 
n_k\,N+n_k'\,N^{-(1+\lambda)/(2+\lambda)}
\end{equation}
with 
\begin{eqnarray*}
n_k&=&(2+\lambda)\,{\Gamma(3+2\lambda)\over\Gamma(1+\lambda)}\,
{\Gamma(k+\lambda)\over\Gamma(k+3+2\lambda)}\,,\\
n_k'&=&-{2+\lambda\over 3+2\lambda}\,
{3^{(3+2\lambda)/(2+\lambda)}\over\Gamma(1+\lambda)}\,
{\Gamma(k+\lambda)\over\Gamma(k)}\,.
\end{eqnarray*}
The corresponding leading behaviors are $n_k\propto k^{-(3+\lambda)}$ and
$n_k'\propto k^{\lambda}$.  Thus the two contributions to the degree
distribution in Eq.~(\ref{NkAw-soln}) are comparable when $k\approx
N^{1/(2+\lambda)}$.  This value coincides with maximal degree $k_{\rm max}$
that is obtained by the extreme value condition $\sum_{k\geq k_{\rm max}}
N/k^{3+\lambda}\approx 1$.  Once again the degree distribution is described
by a scaling function in the dimensionless variable
$\xi=k/N^{1/(2+\lambda)}$.
 
\section{Discrete Approach}

We now turn now to the discrete approach for the network evolution.  That is,
one link is introduced at each discrete time step; this corresponds exactly
to what occurs in the simulation.  We again focus on the caes of the linear
attachment rate $A_k=k$.  We first treat in detail the case of nodes of
degree one and then extend our approach to nodes of higher degrees.  Finally,
we give a scaling description for the degree distribution itself.

\subsection{Nodes of Degree One}

The number of nodes of degree one, $N_1(N)$, is a random variable that
changes according to
\begin{equation}
\label{N1}
N_1(N+1)=\cases{N_1(N)    & prob. ${\displaystyle {N_1\over 2N}}$\cr
                N_1(N) +1 & prob. ${\displaystyle 1-{N_1\over 2N}}$}
\end{equation}
after each node addition event.  That is, with probability $N_1/2N$, a
newly-introduced node attaches to a node of degree one; in this case, the
number of nodes of degree one does not change.  Conversely, with probability
$(1-N_1/2N)$, the new node attaches to a node of degree greater than one and
$N_1$ thus increases by one.  Therefore
\begin{eqnarray*}
\left\langle N_1(N+1)\right\rangle
&=&\left\langle {N_1^2(N)\over 2N}\right\rangle\\
&+&\left\langle N_1(N)+1-{N_1^2(N)\over 2N}-{N_1(N)\over 2N}\right\rangle,
\end{eqnarray*}
from which
\begin{equation}
\label{N1av}
\left\langle N_1(N+1)\right\rangle
=1+\left(1-{1\over 2N}\right)\langle N_1(N)\rangle.
\end{equation}
We take the initial condition $\langle N_1(1)\rangle=N_1(1)=2$.

We solve this recursion in terms of the generating function ${\mathcal
  X}_1(w)=\sum_{N\geq 1} \langle N_1(N)\rangle\, w^{N-1}$.  We therefore
multiply Eq.~(\ref{N1av}) by $Nw^{N-1}$ and sum over $N\geq 1$ to convert
this recursion into the differential equation
\begin{equation}
\label{Xz}
{d {\mathcal X}_1\over dw}={1\over (1-w)^2}+{1\over 2}\,{\mathcal X}_1
+w\,{d {\mathcal X}_1\over dw}\,.
\end{equation}
Solving Eq.~(\ref{Xz}) subject to the initial condition ${\mathcal X}_1(0)=2$
gives
\begin{equation}
\label{Xzsol}
{\mathcal X}_1(w)={2\over 3}\,{1\over (1-w)^2}
+{4\over 3}\,{1\over (1-w)^{1/2}}\,.
\end{equation}
Finally, we expand ${\mathcal X}_1(w)$ in a Taylor series in $w$ to obtain
\begin{equation}
\label{XN}
\langle N_1(N)\rangle={2\over 3}\,N
+{4\over 3\sqrt{\pi}}\,\,{\Gamma\left(N-{1\over 2}\right)\over \Gamma(N)}\,.
\end{equation}
The leading term is identical to that in the continuum approach (cf.~Appendix
A), but the coefficient of the correction term is ${4/(3\sqrt{\pi})}\approx
0.7523$, compared to ${4/3}$ in the continuum approach.

The discrete approach is also suited to analyzing higher moments of the
random variable $N_1(N)$.  The second moment $\langle N_1^2(N)\rangle$ plays
an especially important role as we can then obtain the variance
$\sigma_1^2=\langle N_1^2(N)\rangle-\langle N_1(N)\rangle^2$ and thereby
quantify fluctuations.  From Eq.~(\ref{N1}) this second moment $\langle
N_1^2(N)\rangle$ obeys the following recursion formula
\begin{eqnarray}
\label{N2av}
\left\langle N_1^2(N+1)\right\rangle
&=&1+\left(1-{1\over N}\right)\langle N_1^2(N)\rangle\nonumber\\
&+&\left(2-{1\over 2N}\right)\langle N_1(N)\rangle.
\end{eqnarray}
The solution to this recursion is outlined in Appendix~B and the final result
is
\begin{eqnarray}
\label{N12av} 
\langle N_1^2(N)\rangle&=&{4\over 9}\,N(N+1)-{1\over 3}\,N
+{16\over 9\sqrt{\pi}}\,\,
{\Gamma\left(N+{1\over 2}\right)\over \Gamma(N)}\nonumber\\
&-&{4\over 3\sqrt{\pi}}\,\,{\Gamma\left(N-{1\over 2}\right)\over \Gamma(N)}
 +{35\over 9}\,\delta_{N,1}.
\end{eqnarray}
In the large $N$ limit, we use Stirling's formula to give, for the
variance,
\begin{equation}
\label{sigma1}
\sigma_1^2={N\over 9}-{20\over 9\sqrt{\pi}}\,\,{1\over N^{1/2}}
-{16\over 9\sqrt{\pi}}\,\,{1\over N}+...
\end{equation}

To obtain the entire probability distribution $P(N_1,N)$ one must solve
\begin{eqnarray}
\label{PNN} 
P(N_1,N+1)&=&{N_1\over 2N}\,P(N_1,N)\nonumber\\
&+&\left(1-{N_1-1\over 2N}\right)P(N_1-1,N).
\end{eqnarray}
By the Markov nature of the process, $P(N_1,N)$ should approach a Gaussian
distribution in the large $N$ limit.  Numerically, we indeed find a Gaussian
distribution with a peak at $2N/3$ and dispersion ${1\over 3}\,\sqrt{N}$ in
agreement with our theoretical results for $\langle N_1(N)\rangle$ and
$\langle N_1^2(N)\rangle$.

\subsection{Degree Greater Than One}

For $k\geq 2$, the random variable $N_k\equiv N_k(N)$ changes 
according to
\begin{eqnarray}
\label{Nk-cases}
N_k(N+1)=\cases{
N_k-1   & prob. ${\displaystyle {kN_k\over 2N}}$\cr\cr
N_k+1   & prob. ${\displaystyle {(k-1)N_{k-1}\over 2N}}$\cr\cr
N_k     & prob. ${\displaystyle 1-{(k-1)N_{k-1}+kN_k\over 2N}}$}
\end{eqnarray}
at each node addition event.  Again, because of the Markov nature of this
process, we anticipate that $P(N_k,N)$ approaches a Gaussian distribution for
every {\em fixed} degree $k$; therefore, we only need calculate $\langle
N_k(N)\rangle$ and $\langle N_k^2(N)\rangle$ to infer the asymptotic
distribution.  To determine the first moment, we repeat the steps described
in detail for $k=1$ and obtain the recursion formula
\begin{eqnarray}
\label{Nkav}
\left\langle N_k(N+1)\right\rangle
&=&\left\langle N_k(N)\right\rangle \nonumber\\
&+&\left\langle {(k-1)N_{k-1}(N)-kN_k(N)\over 2N}\right\rangle.
\end{eqnarray}
The solution to this recursion is given in Appendix~C and explicit formulae
for $\langle N_k(N)\rangle$ for $k\leq 5$ are also quoted.  Qualitatively,
these results closely correspond to the asymptotic series for $\langle
N_k(N)\rangle$ in the continuum formulation (Eq.~(\ref{nkn})) but with
somewhat different coefficients in the correction terms.

The determination of the second moment $\langle N_k^2\rangle$ is more
complicated because it is coupled to $\langle N_{k-1}N_k\rangle$, which in
turn is coupled to $\langle N_{k-2}N_k\rangle$, {\it etc}.  However, we can
still determine $\langle N_k^2\rangle$ for small $k$ (Appendix~D).  From the
structure of the rate equations, our general conclusion is that
$\sigma_k^2=\langle N_k^2(N)\rangle-\langle N_k(N)\rangle^2=\mu_k N$.
Therefore the distribution of $N_k(N)$ approaches a Gaussian for each $k$ as
$N\to\infty$.

\subsection{Generating Function Approach}

In close analogy with Sec.~III, we now obtain the generating function for
$\langle N_k(N)\rangle$, from which the exact scaling function in
Eq.~(\ref{Nkscal}) can be deduced.  Since Eq.~(\ref{Nkav}) involves two
discrete variables, $k$ and $N$, it proves useful to introduce the
two-variable generating function
\begin{equation}
\label{Nwz}
{\mathcal N}(w,z)=\sum_{N=1}^\infty \sum_{k=1}^\infty 
\langle N_k(N)\rangle\, w^{N-1}\,z^k\,.
\end{equation}
The governing equation for ${\mathcal N}(w,z)$ that is obtained from
Eq.~(\ref{Nkav}), is
\begin{equation}
\label{gNwz}
\left(2(1-w)\,{\partial\over \partial w}
+z(1-z)\,{\partial\over \partial z}-2\right)
{\mathcal N}={2z\over (1-w)^2}\,.
\end{equation}

This is similar to Eq.~(\ref{gN}) and can be solved accordingly.  We
introduce the rotated variables $x, y$ such that
\begin{equation}
\label{xyw}
x+y=-{1\over 2}\,\ln (1-w), \qquad x-y=\ln{z\over 1-z}, 
\end{equation}
to recast Eq.~(\ref{gNwz}) into
\begin{equation}
\label{gNxyw}
\left({\partial \over \partial x}-2\right){\mathcal N}(x,y)
={2\, e^{5x+4y}\over e^x+e^y}.
\end{equation}
The general solution is
\begin{eqnarray*}
{\mathcal N}(x,y)&=&
e^{4x+4y}-2e^{3x+5y}\\
&+&2e^{2x+6y}\ln\left(e^x+e^y\right)+e^{2x}G(y),
\end{eqnarray*}
and the function $G(y)$ is found from the initial condition ${\mathcal
  N}(w=0,z)=2z$.  When $w=0$, we have $x=-y={1\over 2}\,\ln[z/(1-z0]$, and
hence ${\mathcal N}(-y,y)=2/(1+e^{2y})$.  Therefore
\begin{eqnarray*}
G(y)={2e^{2y}\over 1+e^{2y}}-e^{2y}
+2e^{4y}-2e^{6y}\ln \left(e^{-y}+e^y\right)\,,
\end{eqnarray*}
and finally
\begin{eqnarray}
\label{Nxy} 
{\mathcal N}(x,y)&=&e^{4x+4y}-2e^{3x+5y}-e^{2x+2y}
+2e^{2x+4y}\nonumber\\
&+&2\,{e^{2x+2y}\over 1+e^{2y}}
+2e^{2x+6y}\ln\left({e^{x+y}+e^{2y}\over 1+e^{2y}}\right).
\end{eqnarray}
In term of the original $w, z$ variables, 
\begin{eqnarray}
\label{Nwzsol}
{\mathcal N}(w,z)&=&{(3-2z^{-1})\over (1-w)^{2}}-{1\over 1-w}\nonumber\\
&+&{2(z^{-1}-1)\over (1-w)^{3/2}}
+{2(1-w)^{-1/2}\over (z^{-1}-1)+(1-w)^{1/2}}\nonumber\\
&-&{2(z^{-1}-1)^2\over (1-w)^2}\,\ln\left[1-z+z(1-w)^{1/2}\right].
\end{eqnarray}
By expanding ${\mathcal N}(w,z)$, we can in principle determine all the
$\langle N_k(N)\rangle$.

\subsection{Scaling Function}

To extract the scaling function $F(\xi)$ from the generating function
${\mathcal N}(w,z)$ we use the same approach as in Sec.~IV.  The details are
given in Appendix~E and the final result is
\begin{equation}
\label{Fxisol}
F(\xi)={\rm erfc}\left({\xi\over 2}\right)
+{2\xi+\xi^3\over \sqrt{4\pi}}\,\,e^{-\xi^2/4}\,,
\end{equation}
where erfc$(x)$ is the complementary error function.  A similar result for a
related network model was found previously by Dorogovtsev et al.~\cite{DMS2}.
Notice that the exact form for $F(\xi)$ vanishes much more quickly than
predicted by the continuum approach.  When $k\gg\sqrt{N}$, the continuum
approach gives
\begin{equation}
\label{contNk}
\langle N_k(N)\rangle_{\rm cont.} \to {2\over \sqrt{N}}\,\,\,
e^{-k/\sqrt{N}}\,,
\end{equation}
while the exact average degree distribution has a Gaussian large-degree tail
\begin{equation}
\label{contNk-exact}
\langle N_k(N)\rangle_{\rm exact} \to {2\over \sqrt{\pi N}}\,\,\,
e^{-k^2/4N}\,,
\end{equation}

\begin{figure}
  \narrowtext \hskip 0.0in \epsfxsize=3.0in 
\epsfbox{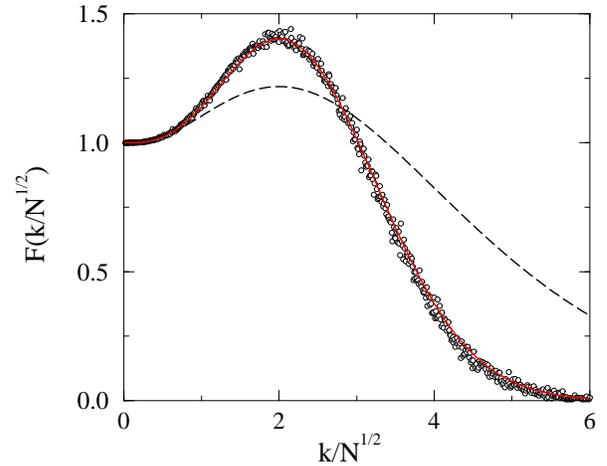} \vskip 0.1in
\caption{Comparison between the scaling function $F(\xi)$, with
  $\xi=k/N^{1/2}$, in the continuum approximation [Eq.~(\ref{Fscal2}), dashed
  curve] and in the discrete approach [Eq.~(\ref{Fxisol}), solid curve].  The
  circles give the simulation data of $10^6$ realizations of a network with
  $N=10^4$ links for the dimer initial condition; these data coincide with
  the theoretical prediction.
\label{compare}}
\end{figure}

The scaling function in Eq.~(\ref{Fxisol}) quantitatively accounts for the
shoulder in the degree distribution.  In contrast, while the scaling function
from the continuum approach does exhibit a peak, it is both quantitatively
and qualitatively inaccurate (Fig.~\ref{compare}).

\section{Higher Moments and Their Fluctuation}

We now turn to higher moments of the degree distribution, as well as the
fluctuation in these quantities between different realizations of the
network.  While the zeroth and first moments of the degree distribution are
simply related to the total number of links for {\em any} network topology,
the higher moments are not so simply characterized, but instead reflect the
power-law tail of the degree distribution.

We first compare the moments of the average degree distribution to appreciate
the difference between the continuum and exact descriptions.  For the second
moment, we use the identity
\begin{equation}
\label{second}
\sum_{k=1}^\infty k(k+1)\langle N_k\rangle\equiv 
\left(z^2\,{\partial\over \partial z}\right)^2
{\mathcal N}(N,z)\Big|_{z=1}.
\end{equation}
Using ${\mathcal N}(N,z)$ from Eq.~(\ref{exact}), together with the value of
the first moment, we obtain, in the continuum approximation,
\begin{equation}
\label{2cont}
\langle k^2\rangle_{\rm cont.} \equiv 
\sum_{k=1}^\infty k^2\langle N_k\rangle_{\rm cont.}=
2N\ln N + 2N.
\end{equation}
On the other hand, using the exact discrete expression (\ref{Nwzsol}) we find
\begin{eqnarray*}
\left(z^2\,{\partial\over \partial z}\right)^2
{\mathcal N}(w,z)\Big|_{z=1}=&&{4-2\ln(1-w)\over (1-w)^2}\,,
\end{eqnarray*}
which we then expand in a series in $w$ to yield, for the second moment,
\begin{equation}
\label{2exact}
\langle k^2\rangle_{\rm exact}\equiv 
\sum_{k=1}^\infty k^2\langle N_k\rangle_{\rm exact}=2N H_N\,.
\end{equation}
Here $H_N=\sum_{1\leq j\leq N} j^{-1}$ is the harmonic number\cite{knuth}. 
In the large $N$ limit, therefore,
\begin{eqnarray*}
\langle k^2\rangle_{\rm exact}=2N \ln N + 2\gamma N+1-{1\over 6N}+\ldots\,,
\end{eqnarray*}
where $\gamma\cong 0.5772166$ is Euler's constant.  

For higher moments, even the leading term given by the continuum approach is
erroneous.  For example, 
\begin{equation}
\label{3cont}
\langle k^3\rangle_{\rm cont.}= 24N^{3/2}-6N\ln N - 22N\,,
\end{equation}
while the exact value is
\begin{equation}
\label{3exact}
\langle k^3\rangle_{\rm exact}=
{32\over \sqrt{\pi}}\,{\Gamma\left(N+{3\over 2}\right)\over\Gamma(N)}
-6N H_N-16N\,.
\end{equation}

More generally, the dependence of the moments on $N$ stems from the power-law
tail of the degree distribution $\langle N_k\rangle\propto N/k^3$.  From this
asymptotic distribution, a suitably normalized set of measures for the mean
degree 
\begin{equation}
\label{Mn}
{\mathcal M}_n=\left({\langle k^n\rangle\over{\langle
      k^0\rangle}}\right)^{1/n},
\end{equation}
has the following $N$ dependence:
\begin{equation}
\label{Mn-cases}
{\mathcal M}_n\propto \cases{{\rm const.} & $n<2$ \cr
                                    \ln N & $n=2$ \cr
                              N^{(n-2)/2} & $n>2$ }
\end{equation}

\begin{figure}
  \narrowtext \hskip 0.0in \epsfxsize=2.8in 
\epsfbox{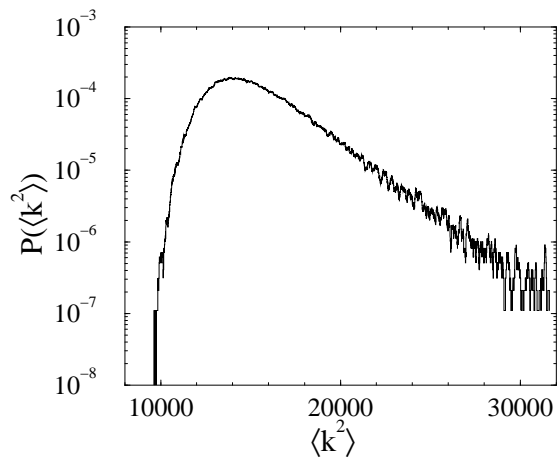} \vskip 0.1in
\caption{Distribution of $\langle k^2\rangle$ for $10^5$ realizations of a
  growing network with $N=10^3$ for attachment rate $A_k=k$ with the triangle
  initial condition.  The raw data has been smoothed over a 100-point range.
\label{Pksq}}
\end{figure}

In a related vein, we also study the fluctuations in these moments between
different realizations of the network growth.  That is, we record the value
of $\langle k^2\rangle$ for each realization of the network to obtain the
underlying distribution $P(\langle k^2\rangle)$.  A typical result is shown
in Fig.~\ref{Pksq}.  Notice that the distribution of $\langle k^2\rangle$ is
relatively broad with an exponential tail.  The distributions of higher
moments are even broader, with each being dominated by the realizations with
the largest value of the corresponding moment.

\section{Concluding Remarks}

We studied the role of finiteness on the degree distributions of growing
networks with a node attachment rate of the form $A_k=k+\lambda$.  For finite
networks, fluctuations are no longer negligible and a stochastic approach is
needed to analyze these properties.  We found the average degree distribution
within an approximate continuum formulation and by an exact discrete
approach.  The continuum approach has the advantage of being much simpler
than the discrete formulation, but does not provide a quantitatively accurate
description of the large-$k$ tail of the degree distribution.

We also argued that the degree distribution $N_k(N)$, when considered as the
random variable in $k$, exhibits self averaging, {\it i.e.}, the relative
fluctuations in $N_k(N)$ diminish as $N\to\infty$.  Moreover, the variance
$\sigma_k^2=\langle N_k^2\rangle-\langle N_k\rangle^2$ varies linearly with
$N$, and the probability distribution $P(N_k,N)$ approaches a Gaussian.  To
support these assertions, we computed $\sigma_k^2$ for $k=1,2$.  These
partial results support our general hypothesis that fluctuations in $N_k(N)$
are Gaussian.  Perhaps the Van Kampen $\Omega$-expansion \cite{Van} would
prove to be a more appropriate analysis tool to undertake a systematic study
of fluctuations in growing networks.

Of course, the random variables $N_k(N)$ should be Gaussian only for
sufficiently small $k$, {\it viz.}, as long as $\langle N_k\rangle\gg 1$, or
equivalently, $k\ll N^{1/(3+\lambda)}$.  On the other hand, fluctuations
become large and non-Gaussian when $k\propto N^{1/(3+\lambda)}$.  Determining
the fluctuations in this degree range seems to be difficult, as one must study
the master equation for the joint probability distribution.

In this work, we limited ourselves to the degree distribution; this is
perhaps the most important and also the most easily analyzable local
structural characteristic of a network.  However, recent investigations of
growing networks has increasingly focused on global characteristics, such as
the size distribution of connected components, see {\it e.g.},
Refs.~\cite{clusters,sam,KKKR,bernard}.  The methods described in this paper
should be applicable to probing fluctuations of the component size
distribution and other global network characteristics.  This direction seems
especially exciting since the simplest growing network models that allow for
a multiplicity of clusters exhibit a very unusual infinite-order percolation
transition\cite{clusters,sam,KKKR,bernard}.  Thus one might anticipate
interesting giant fluctuations near the percolation transition of these
models.

We are grateful to NSF grant DMR9978902 for partial financial support of this
research.

\appendix
\section{The average degree distribution in the continuum formulation}

Within the continuum framework, the average degree distribution is described
by Eqs.~(\ref{NkN}).  Successively solving these equations by elementary
methods, we obtain $\langle N_k(N)\rangle$.  For $k=1,2,3$, and 4 we obtain:
\begin{eqnarray*}
\left\langle N_1(N)\right\rangle
&=&{2\over 3}\,N+{4\over 3}\,{1\over N^{1/2}},\\
\left\langle N_2(N)\right\rangle
&=&{1\over 6}\,N+{4\over 3}\,{1\over N^{1/2}}-{3\over 2}\,{1\over N},\\
\left\langle N_3(N)\right\rangle
&=&{1\over 15}\,N+{4\over 3}\,{1\over N^{1/2}}-3\,{1\over N}
+{8\over 5}\,{1\over N^{3/2}},\\
\left\langle N_4(N)\right\rangle
&=&{1\over 30}\,N+{4\over 3}\,{1\over N^{1/2}}-{9\over 2}\,{1\over N}
+{24\over 5}\,{1\over N^{3/2}}-{5\over 3}\,{1\over N^{2}}.
\end{eqnarray*}

\section{Generating Function for $\langle N_1^2(N)\rangle$}

To determine $\langle N_1^2(N)\rangle$, we introduce the generating function
${\mathcal Y}_1(w)=\sum_{N\geq 1} \langle N_1^2(N)\rangle\, w^{N-1}$.
This converts the recursion relation Eq.~(\ref{N2av}) into the differential
equation for the generating function
\begin{equation}
\label{Yz}
(1-w)\,{d {\mathcal Y}_1\over dw}={1\over (1-w)^2}+{3\over 2}\,{\mathcal X}_1
+2w\,{d {\mathcal X}_1\over dw}\,,
\end{equation}
with ${\mathcal X}_1(w)$ given by Eq.~(\ref{Xzsol}).  Solving (\ref{Yz})
subject to the initial condition ${\mathcal Y}_1(0)=4$ we obtain
\begin{eqnarray}
\label{Yw}
{\mathcal Y}_1(w)&=&{8\over 9}\,{1\over (1-w)^3}
-{1\over 3}\,{1\over (1-w)^2}+{8\over 9}\,{1\over (1-w)^{3/2}}\nonumber\\
&-&{4\over 3}\,{1\over (1-w)^{1/2}}+{35\over 9}\,,
\end{eqnarray}
Expanding this generating function in a Taylor series then yields the result
for $\langle N_1^2(N)\rangle$ quoted in Eq.~(\ref{N12av}).

\section{Generating function for first moment}

Here we solve the recursion formula Eq.~(\ref{Nkav}) for $\langle
N_k(N)\rangle$.  We first introduce the generating function ${\mathcal
  X}_k(w)=\sum_{N=1}^\infty \langle N_k(N)\rangle\, w^{N-1}$ to eliminate the
variable $N$ and convert Eq.~(\ref{Nkav}) into a differential equation that
relates ${\mathcal X}_k$ and ${\mathcal X}_{k-1}$.  This equation is further
simplified by making the transformation
\begin{equation}
\label{XwUu}
{\mathcal X}_k(w)=(1-w)^{{k\over 2}-1}\,{\mathcal U}_k(u), \quad
u={1\over \sqrt{1-w}}-1.
\end{equation}
The resulting equation is  
\begin{equation}
\label{Uu}
{d {\mathcal U}_k\over du}=(k-1)\,{\mathcal U}_{k-1}, \qquad k\geq 2.
\end{equation}
Rewriting our previous solution (\ref{Xzsol}) as 
\begin{equation}
\label{U1}
{\mathcal U}_1(u)={2\over 3}\,u^3+2u^2+2u+2,
\end{equation}
one can solve Eqs.~(\ref{Uu}) subject to the initial condition ${\mathcal
  U}_k(u=0)=0$ for $k\geq 2$.  The final result is
\begin{eqnarray*}
{\mathcal U}_k(u)={4 u^{k+2}\over k(k+1)(k+2)}
+{4 u^{k+1}\over k(k+1)}
+{2 u^k\over k}+2u^{k-1}\,.
\end{eqnarray*}
Using the binomial formula, we transform ${\mathcal X}_k(z)$ into the series
\begin{eqnarray*}
{\mathcal X}_k(w)&=&{4\over k(k+1)(k+2)}\,{1\over (1-w)^2}
+{4\over 3}\,{1\over (1-w)^{1/2}}\\
&+&2\sum_{a=1}^{k-1} (-1)^a\,{a+2\over a+3}\,{k-1\choose a}\,(1-w)^{(a-1)/2}\,.
\end{eqnarray*}
Expanding ${\mathcal X}_k(w)$ in a Taylor series in $w$ we obtain
$\left\langle N_k(N)\right\rangle$.  The analytic expressions for
$\left\langle N_k(N)\right\rangle$ with $k\leq 5$ are obtained by expanding
${\mathcal X}_k(w)$ in a Taylor series.  This gives
\begin{eqnarray*}
\left\langle N_1(N)\right\rangle
&=&{2\over 3}\,N+{4\over 3\sqrt{\pi}}\,\,
{\Gamma\left(N-{1\over 2}\right)\over \Gamma(N)},\\
\left\langle N_2(N)\right\rangle
&=&{1\over 6}\,N+{4\over 3\sqrt{\pi}}\,\,
{\Gamma\left(N-{1\over 2}\right)\over \Gamma(N)}-{3\over 2}\,\delta_{N,1},\\
\left\langle N_3(N)\right\rangle&=&{1\over 15}\,N
+{4\over 3\sqrt{\pi}}\,\,{\Gamma\left(N-{1\over 2}\right)\over \Gamma(N)}
-3\,\delta_{N,1}\\
&-&{4\over 5\sqrt{\pi}}\,\,{\Gamma\left(N-{3\over 2}\right)\over \Gamma(N)},\\
\left\langle N_4(N)\right\rangle&=&{1\over 30}\,N
+{4\over 3\sqrt{\pi}}\,\,{\Gamma\left(N-{1\over 2}\right)\over \Gamma(N)}
-{9\over 2}\,\delta_{N,1}\\
&-&{12\over 5\sqrt{\pi}}\,\,{\Gamma\left(N-{3\over 2}\right)\over \Gamma(N)}
-{5\over 3}\,\delta_{N,1}+{5\over 3}\,\delta_{N,2}\,\\
\left\langle N_5(N)\right\rangle&=&{2\over 105}\,N
+{4\over 3\sqrt{\pi}}\,\,{\Gamma\left(N-{1\over 2}\right)\over \Gamma(N)}
-6\,\delta_{N,1}\\
&-&{24\over 5\sqrt{\pi}}\,\,{\Gamma\left(N-{3\over 2}\right)\over \Gamma(N)}
-{20\over 3}\,\delta_{N,1}+{20\over 3}\,\delta_{N,2}\,\\
&+&{9\over 7\sqrt{\pi}}\,\,{\Gamma\left(N-{5\over 2}\right)\over \Gamma(N)}\,.
\end{eqnarray*}

Generally, there are slightly different formulae for even
\begin{eqnarray*}
\left\langle N_{2k}(N)\right\rangle&=&n_{2k}\,N
+\sum_{j=1}^k A_{kj}\,\delta_{Nj}\\
&+&\sum_{i=0}^{k-1}{4(i+1)\over 2i+3}\,{2k-1\choose 2i}\,
{\Gamma\left(N-{1\over 2}-i\right)\over 
\Gamma\left({1\over 2}-i\right)\,\Gamma(N)}
\end{eqnarray*}
and odd 
\begin{eqnarray*}
\left\langle N_{2k+1}(N)\right\rangle&=&n_{2k+1}\,N
+\sum_{j=1}^k B_{kj}\,\delta_{Nj}\\
&+&\sum_{i=0}^{k}{4(i+1)\over 2i+3}\,{2k\choose 2i}\,
{\Gamma\left(N-{1\over 2}-i\right)\over 
\Gamma\left({1\over 2}-i\right)\,\Gamma(N)}
\end{eqnarray*}
indices.  Here the $n_k$ are given by Eqs.~(\ref{nk}) and explicit
expressions for the coefficients $A_{kj}$ and $B_{kj}$ could be found by
expanding the polynomials in the generating functions ${\mathcal X}_{2k}(w)$
and ${\mathcal X}_{2k+1}(w)$.

\section{Higher Moments}

Starting from Eq.~(\ref{Nk-cases}), a straightforward computation yields
\begin{eqnarray}
\label{Nk2}
\left\langle N_k^2\right\rangle
&=&\left(1-{k\over N}\right)\left\langle N_k^2\right\rangle 
+{k-1\over N}\left\langle N_{k-1}N_k\right\rangle\nonumber\\
&+&\left\langle {(k-1)N_{k-1}+kN_k\over 2N}\right\rangle\,,
\end{eqnarray}
where the correlation function on the left-hand side is a function of $N+1$
and those on the right-hand side are functions of $N$.  Obviously,
$\left\langle N_k^2\right\rangle$ is coupled with $\left\langle
  N_{k-1}N_k\right\rangle$.  The recursion relation for this correlation
function reads (for $k\geq 3$)
\begin{eqnarray}
\label{Nk3}
\left\langle N_{k-1}N_k\right\rangle
&=&\left(1-{2k-1\over 2N}\right)\left\langle N_{k-1}N_k\right\rangle
+{k-1\over 2N}\left\langle N_{k-1}^2\right\rangle\nonumber\\
&+&{k-2\over 2N}\left\langle N_{k-2}N_k\right\rangle
-{k-1\over 2N}\langle N_{k-1}\rangle.
\end{eqnarray}

Fortunately no higher-order correlation functions appear, and additionally
the total index decreases, {\it i.e.}, $\left\langle N_k^2\right\rangle$,
whose total index is $2k$, involves the correlation function $\left\langle
  N_{k-1}N_k\right\rangle$, whose total index is $2k-1$.  One therefore can
determine all correlation functions by starting from the smallest total index
and then working up to larger indices.  For example, the first non-trivial
correlation function $\left\langle N_1N_2\right\rangle$ whose total index
equals three satisfies an equation slightly different from the general form
of Eq.~(\ref{Nk3}), {\it viz.},
\begin{eqnarray}
\label{Nk4}
\left\langle N_1N_2\right\rangle
&=&\left(1-{3\over 2N}\right)\left\langle N_1N_2\right\rangle\nonumber\\
&+&{1\over 2N}\left\langle N_1^2\right\rangle
+\left(1-{1\over N}\right)\left\langle N_2\right\rangle\,.
\end{eqnarray}
Notice here that we already know $\left\langle N_1^2\right\rangle$.

We can solve for $\left\langle N_1N_2\right\rangle$ using the generating
function technique.  We define the generating function ${\mathcal
  Z}_1(w)=\sum_{N\geq 1} \langle N_1(N)N_2(N)\rangle\, w^{N-1}$ which
satisfies the differential equation
\begin{equation}
\label{Zw}
2(1-w)\,{d {\mathcal Z}_1\over dw}=-{\mathcal Z}_1
+{\mathcal Y}_1\,+2w\,{d {\mathcal X}_2\over dw},
\end{equation}
with solution
\begin{eqnarray}
\label{Zwsol}
{\mathcal Z}_1(w)&=&{2\over 9}\,{1\over (1-w)^3}
-{1\over 5}\,{1\over (1-w)^2}+{5\over 9}\,{1\over (1-w)^{3/2}}\nonumber\\
&-&{4\over 3}\,{1\over (1-w)^{1/2}}-{47\over 15}\,(1-w)^{1/2}+{35\over 9}\,.
\end{eqnarray}
Expanding ${\mathcal Z}_1(w)$ in a power series in $w$ we obtain
\begin{eqnarray*}
\langle N_1N_2\rangle&=&{1\over 9}\,N(N+1)-{1\over 5}\,N
+{10\over 9\sqrt{\pi}}\,\,
{\Gamma\left(N+{1\over 2}\right)\over \Gamma(N)}\\
&-&{4\over 3\sqrt{\pi}}\,\,{\Gamma\left(N-{1\over 2}\right)\over \Gamma(N)}
+{47\over 30\sqrt{\pi}}\,\,{\Gamma\left(N-{3\over 2}\right)\over \Gamma(N)}\\
&+&{35\over 9}\,\delta_{N,1}\,. 
\end{eqnarray*}
Asymptotically, $\langle N_1N_2\rangle\to \langle N_1\rangle \langle
N_2\rangle$ as expected.

There are two correlation functions, $\left\langle N_2^2\right\rangle$ and
$\left\langle N_1N_3\right\rangle$, whose total index equals four.  The
former satisfies Eq.~(\ref{Nk2}) with $k=2$, {\it i.e.},
\begin{eqnarray*}
\left\langle N_2^2\right\rangle
=\left(1-{2\over N}\right)\left\langle N_2^2\right\rangle 
+\left\langle {N_{1}+2N_2+2N_1N_2\over 2N}\right\rangle\,,
\end{eqnarray*}
from which we determine the generating function
\begin{eqnarray*}
{\mathcal Y}_2(w)&=&{1\over 18}\,{1\over (1-w)^3}
+{1\over 10}\,{1\over (1-w)^2}+{2\over 9}\,{1\over (1-w)^{3/2}}\\
&+&{4\over 9}\,{1\over (1-w)^{1/2}}
-{94\over 15}\,(1-w)^{1/2}+{49\over 9}-{55\over 18}\,w\,.
\end{eqnarray*}
Expanding ${\mathcal Y}_2(w)$ we obtain 
\begin{eqnarray*}
\langle N_2^2(N)\rangle&=&{1\over 36}\,N(N+1)+{1\over 10}\,N
+{4\over 9\sqrt{\pi}}\,\,
{\Gamma\left(N+{1\over 2}\right)\over \Gamma(N)}\\
&+&{4\over 9\sqrt{\pi}}\,\,{\Gamma\left(N-{1\over 2}\right)\over \Gamma(N)}
+{47\over 15\sqrt{\pi}}\,\,{\Gamma\left(N-{3\over 2}\right)\over \Gamma(N)}\\
&+&{49\over 9}\,\delta_{N,1}-{55\over 18}\,\delta_{N,2}\,. 
\end{eqnarray*}
In the large $N$ limit, we find that variance grows linearly with $N$
according to $\sigma_2^2\sim{23\over 180}\,N$.  It appears that
\begin{equation}
\label{sigmak}
\sigma_k^2\to \mu_k\,N \qquad {\rm as} \quad N\to\infty,
\end{equation}
for all $k$, although we solved only the cases $k=1$ and 2, where
$\mu_1={1\over 9}$ and $\mu_2={23\over 180}$.

\section{Scaling Function in the Discrete Approach}

To extract the scaling function from the generating function ${\mathcal
  N}(w,z)$ we adapt the technique employed in Sec.~IV for discrete variables.
We first write
\begin{equation}
\label{sw}
z^{-1}=1+s\,\sqrt{1-w}
\end{equation}
and keep $s$ finite while taking the $w\to 1$ limit.  We again consider the
modified generating function
\begin{equation}
\label{modN}
\left(z^2\,{\partial \over \partial z}\right)^3{\mathcal N}
=\sum_{N=1}^\infty \sum_{k=1}^\infty 
4 N F\left({k\over\sqrt{N}}\right)\, w^{N-1} z^{k+3}\,.
\end{equation}
On the right-hand side of this equation we have already replaced 
$(k+2)(k+1)k\langle N_k(N)\rangle$ by $4 N F(k/\sqrt{N})$ as implied by    
Eqs.~(\ref{nk})--(\ref{Nkscal}).

Substituting the exact expression (\ref{Nwzsol}) for the generating function
into the left-hand side of Eq.~(\ref{modN}) and keeping only the dominant
contribution gives
\begin{equation}
\label{LHS} 
4(1-w)^{-5/2}J(s),
\end{equation}
with $J(s)$ given by Eq.~(\ref{Js}).  To simplify the right-hand side of
Eq.~(\ref{modN}) we substitute Eq.~(\ref{sw}) and replace the sums by
integrals.  The dominant contribution in the $w\to 1$ limit is
\begin{equation}
\label{RHS} 
4(1-w)^{-5/2}\int_0^\infty d\xi\,e^{-\xi s}\int_0^\infty d\eta\,
\eta\,F(\xi\eta^{-1/2})\,e^{-\eta}\,, 
\end{equation}
where $\xi=k\sqrt{1-w}$ and $\eta=N(1-w)$.  Therefore the double integral in
Eq.~(\ref{RHS}) is equal to $J(s)$.  The double integral can be interpreted
as the Laplace transform $\hat \Phi(s)=\int_0^\infty
d\xi\,\exp(-s\xi)\Phi(\xi)\,$ of the function
\begin{equation}
\label{Phi} 
\Phi(\xi)=\int_0^\infty d\eta\,
\eta\,F(\xi\eta^{-1/2})\,e^{-\eta}\,.
\end{equation}
We already know how to solve $\hat \Phi(s)=J(s)$, so
\begin{equation}
\label{Phisol}
\Phi(\xi)=\left(1+\xi\right)\left(1+{\xi^2\over 2}\right)\,e^{-\xi}\,.
\end{equation}

To determine $F(\xi)$, we must solve the integral equation (\ref{Phisol})
with $\Phi(\xi)$ given by Eq.~(\ref{Phi}).  To solve this integral equation,
notice that $\Phi(\xi)$ is almost a Laplace transform of function $F$. 
Indeed, if instead of $\eta$ and $F(\xi\eta^{-1/2})$ we use $\zeta$ and 
$G(\zeta)$ defined according to   
\begin{equation}
\label{zetaG} 
\zeta={\eta\over \xi^2}, \qquad 
G(\zeta)=\zeta\,F(\zeta^{-1/2})\,,
\end{equation}
then we obtain $\Phi(\xi)=p^2\hat G(p)$, with $p=\xi^2$ being the Laplace
variable and $\hat G(p)=\int_0^\infty d\zeta\,G(\zeta)\,\exp(-p\zeta)$.
Re-writing the integral equation (\ref{Phisol}) in terms of $p$ gives
\begin{eqnarray*}
\hat G(p)=
\left(p^{-2}+p^{-3/2}+{1\over 2}\,p^{-1}+{1\over 2}\,p^{-1/2}\right)
\,\exp(-\sqrt{p})\,.
\end{eqnarray*}
Inverting this Laplace transform yields \cite{AS}
\begin{equation}
\label{Gsol}
G(\zeta)=\zeta\,{\rm erfc}\left({1\over\sqrt{4\zeta}}\right)
+{2\zeta +1\over \sqrt{4\pi\zeta}}\,\,
e^{-1/4\zeta}\,,
\end{equation}
where ${\rm erfc}(x)={2\over \sqrt{\pi}}\int_x^\infty dt\,\exp(-t^2)$ is the
complementary error function.  Since $F(\xi)=\xi^2 G(\xi^{-2})$, see
Eq.~(\ref{zetaG}), we arrive at the scaled average degree distribution quoted
in Eq.~(\ref{Fxisol}).

\end{multicols}
\end{document}